\documentclass[sigconf,9pt]{acmart}
\usepackage{multirow} 
\usepackage{colortbl}
\usepackage{makecell}
\definecolor{mycolor}{rgb}{0.92,0.87, 0.96}

\copyrightyear{2026}
\acmYear{2026}
\setcopyright{cc}
\setcctype{by}
\acmConference[DAC '26]{63rd ACM/IEEE Design Automation Conference}{July 26--29, 2026}{Long Beach, CA, USA}
\acmBooktitle{63rd ACM/IEEE Design Automation Conference (DAC '26), July 26--29, 2026, Long Beach, CA, USA}
\acmDOI{10.1145/3770743.3804056}
\acmISBN{979-8-4007-2254-7/2026/07}

\begin{document}

\title{RTL-Sequencer: Towards Scalable RTL Timing Prediction with the Sequence-based Paradigm}







\author{Ziyan Guo,\ Wenji Fang,\ Wenkai Li,\ Yuchao Wu,\ Shang Liu,\ Zhiyao Xie}
\authornote{Corresponding Author}
\affiliation{%
  \institution{Hong Kong University of Science and Technology (HKUST)}
  \country{\{zguoby,  wfang838, wlidm, ywu092, sliudx\}@connect.ust.hk, eezhiyao@ust.hk}
}

\begin{abstract}
Accurate timing prediction at the register-transfer level (RTL) is a longstanding challenge in design automation. Existing graph-based methods struggle with limited receptive fields, high complexity, and a lack of signal directionality. We present RTL-Sequencer, a novel sequence-based paradigm that enables scalable RTL timing prediction via linearizing logic cones by breadth-first traversal and applying modern linear sequence models. Furthermore, sequence models are customized by four synergistic techniques, including sequence shuffling, bidirectional modeling, differentiable modeling, and a hybrid graph-sequence architecture. Extensive experiments demonstrate significant improvements of RTL-Sequencer over state-of-the-art baselines, advancing early-stage timing optimization.
\end{abstract}

\maketitle

\section{INTRODUCTION}
Timing remains a critical optimization objective in the design of digital integrated circuits. A fundamental limitation in the conventional design methodologies is that accurate timing analysis is only available after the post-layout stage, due to the lack of accurate physical information like wire length information \cite{huang2015opentimer}. Consequently, this delayed feedback undermines the efficacy of timing-driven optimizations during early design phases, including register-transfer level (RTL) design, thereby significantly extending design iterations and increasing time-to-market.

To bridge this gap, Machine Learning (ML) has emerged as a promising data-driven paradigm for early-stage timing prediction. As shown in Table~\ref{tab:summary}, prior research has extensively investigated ML-based timing prediction across various design stages, including the layout, the netlist, and the most challenging RTL stage due to its high level of abstraction \cite{fang2023masterrtl,fang2024annotating,wangbridging}. At the RTL stage, Hardware Description Language (HDL) code is typically transformed into an intermediate representation called the Boolean Operator Graph (BOG) \cite{fang2024annotating,wangbridging}, which is a directed graph of Boolean logic gates. ML–based timing prediction is then applied to logic cones, defined as subgraphs of the BOG that encompass an output register together with its all fan-in input registers and combinational nodes. 

\begin{figure}[t]
    \centering
    \includegraphics[width=\linewidth]{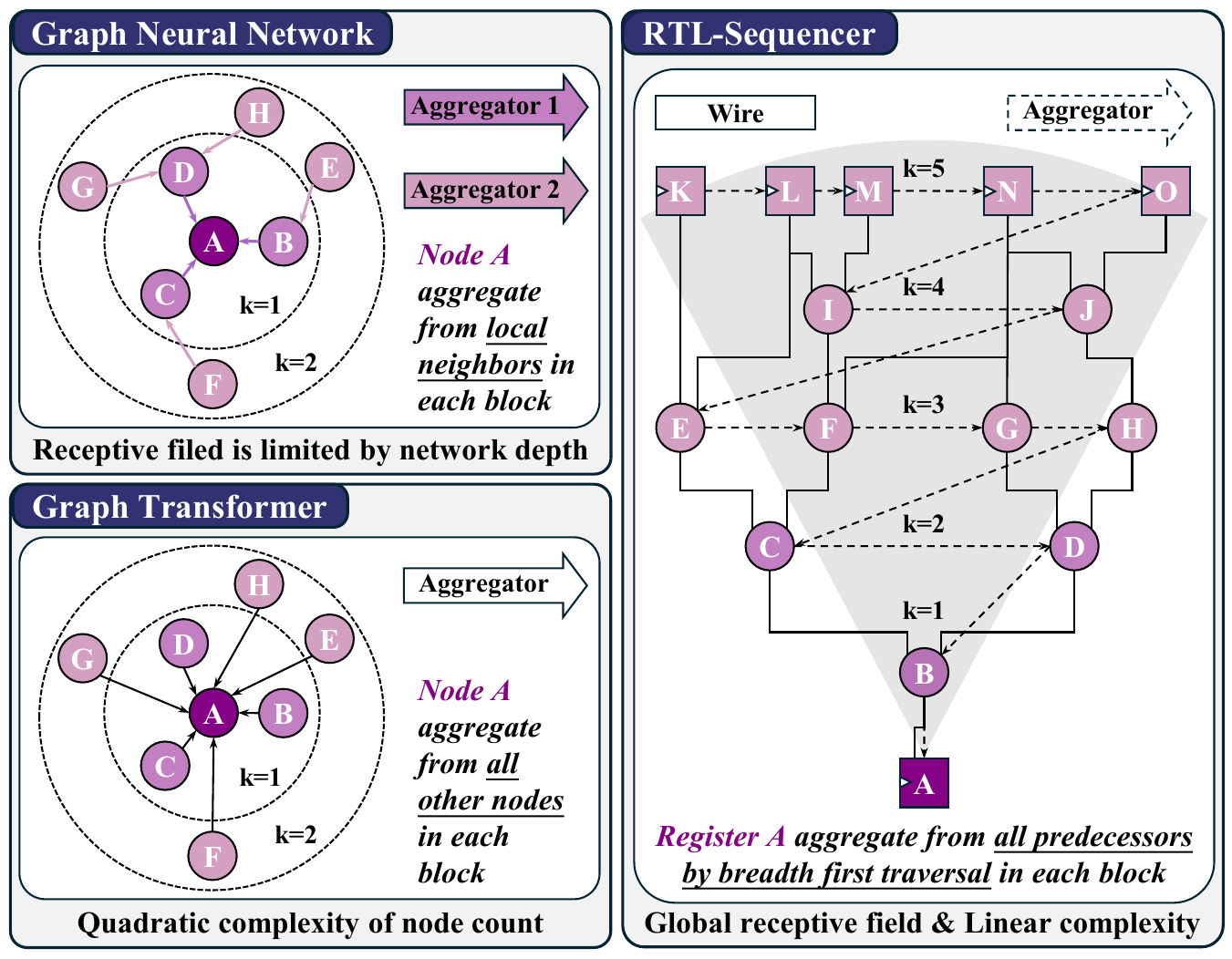}
    \vspace{-7mm}
    \caption{RTL-Sequencer vs. conventional graph-based approaches, such as the Graph Neural Network and the Graph Transformer, highlighting the scalability of the sequence-based paradigm on the receptive field and the node capacity.}
    \label{fig:teaser}
    \vspace{-5.5mm}
\end{figure}

However, we observe that these cones are generated before synthesis and consequently exhibit deep, large-scale topological structures, often comprising more than ten thousand nodes with maximum depths exceeding one hundred. Such complexity presents substantial challenges for ML-based methodologies to accurately model long-range timing dependencies within the cones. As summarized in Table~\ref{tab:summary}, these methodologies can be broadly classified into four categories:
\textbf{(i) Traditional ML}, such as Random Forest \cite{breiman2001random} and XGBoost \cite{chen2016xgboost}, primarily model statistical characteristics of logic cones, thereby neglecting their topological structure that is critical for accurate timing analysis. 
\textbf{(ii) GNNs including Graph Convolutional Networks (GCNs) \cite{kipf2016semi} and Graph Attention Networks (GAT) \cite{velivckovic2017graph}} suffer from constrained receptive fields, which are inherently bounded by network depth \cite{10.1145/3358695.3360934, guo2022timing}. Such restrictions hinder their applicability to pre-synthesis cones with substantial logic depth. Attempts to mitigate this issue by simply stacking additional GNN layers often exacerbate the problem of over-smoothing \cite{rusch2023survey}, resulting in homogenized node embeddings that lack discriminative representation. Consequently, the generalization capacity of these models deteriorates significantly when applied to large-scale designs. 
\textbf{(iii) GNN-Propagation} or DAGNN \cite{thost2021directed} aims to provide a global receptive field by computing representations layer by layer within the cone. Nevertheless, this architecture imposes strict inter-layer dependencies, resulting in training quadratic complexity with cone depth. Such complexity not only impedes training parallelization but also exacerbates error accumulation in deeper layers, thereby constraining scalability to deeply structured designs.
\textbf{(iv) Graph Transformers (GTs)} \cite{yun2019graph} offer global receptive fields; however, they suffer from quadratic complexity with node count and over-emphasize node relationships rather than signal directionality. In addition, TF-Predictor~\cite{cao2022tf} incorporates a Transformer as a sequence model, but it only represents specified paths at the relatively tractable layout stage and incurs quadratic node complexity.
\pagestyle{empty}
\begin{table}[!t]
\centering
\begin{tabular}{ccccc}
\hline
Stage & Type & Works & \makecell{Global \\ Recept.} & \makecell{Linear \\ Complex.}   \\ 
\hline
\multirow{3}{*}{Layout} & Traditional & \cite{6681682,7171706,8807063,9256755,10.1145/3489517.3530598} & $-$ & $-$ \\ 
 & GNN-Prop. & \cite{10.1145/3489517.3530597,10247802,10.1145/3649329.3656251,jin2025crosstalk, 10.1145/3626959} & $\checkmark$ & $\times$   \\ 
 & GT & \cite{9927326,10.1609/aaai.v38i15.29653} & $\checkmark$ & $\times$  \\ 
\hline
\multirow{2}{*}{Netlist} & Traditional & \cite{10.1145/3613424.3623794} & $-$ & $-$ \\
 & GNN & \cite{xie2021net2,9707500}  & $\times$ & $\checkmark$  \\ 
\hline
\multirow{5}{*}{RTL} & Traditional & \cite{10069028,fang2023masterrtl,fang2024annotating} & $-$ & $-$ \\
 & GNN & \cite{10.1145/3508352.3561095} & $\times$ & $\checkmark$ \\
 & GNN-Prop. & \cite{wangbridging,wang2025pre} & $\checkmark$ & $\times$ \\
 & GT & \cite{10.1145/3470496.3527444,10.1145/3658617.3697597,fang2025circuitfusion} & $\checkmark$ & $\times$ \\
 & \cellcolor{mycolor} Sequencer & \cellcolor{mycolor} Ours & \cellcolor{mycolor} $\checkmark$ & \cellcolor{mycolor} $\checkmark$ \\
\hline
\end{tabular}
\vspace{1mm}
\caption{Summary of existing methods for timing prediction.}
\label{tab:summary}
\vspace{-10mm}
\end{table}

To address the inherent limitations of graph-based methodologies, we introduce a novel framework, \textbf{RTL-Sequencer}, which establishes a scalable sequence-based paradigm for the challenging RTL timing prediction. \textbf{Our key insight} lies in customizing linear sequence models \cite{gu2024mamba,dao2024transformers,sun2023retentive,peng2023rwkv,yang2024parallelizing} instead of graph-based models. 
As shown in Fig.~\ref{fig:teaser}, such a paradigm shift is able to offer several \textbf{critical advantages over graph-based models:}
\textbf{(i)} Global receptive fields for long-range timing dependencies independent of network depth;
\textbf{(ii)} Linear computational complexity with respect to node count and cone depth;
\textbf{(iii)} Explicit preservation of signal directionality, thus mitigating over-smoothing; 
\textbf{(iv)} Elimination of strict intra-layer dependencies. 
To achieve this sequence-based paradigm, an inverse Breadth-First Traversal (BFT) from the output register provides a basic solution to convert cones into sequences. 

However, the straightforward adoption of deterministic BFT and conventional sequence models can be sub-optimal, primarily due to the loss of topological information when topology-invariant cones are reduced to fixed, unidirectional, breadth-first node sequences. To address these limitations, we propose \textbf{four synergistic techniques to customize sequence models further} to this challenging task:
\textbf{(i) Sequence Shuffling} to expose sequence models to diverse sequence variants of cones rather than the fixed node ordering introduced by deterministic BFT. Specifically, nodes within each BFT depth are randomly permuted during training as data augmentation, while preserving logical dependencies and timing closure, thereby improving generalization across heterogeneous RTL designs.
\textbf{(ii) Bidirectional Sequence Modeling} to capture the bi-directional flow of signal propagation in logic cones, therefore enabling deeper network blocks, via the representation awareness of later nodes to the earlier nodes.
\textbf{(iii) Differentiable Sequence Modeling} to ensure that node representations align with depth-first critical timing paths rather than breadth-first neighbors. Concretely, we incorporate a differential (i.e., global minus depth-local) mechanism, which extracts depth-specific embeddings via another sequence model and subtracts them from global embeddings, emphasizing global timing behavior rather than depth-local influences.
\textbf{(iv) Hybrid Graph-Sequence Architecture} to harness the complementary strengths of graph-based and sequence-based paradigms in the local breadth of nodes and the global depth of timing paths, respectively. Specifically, we propose an architecture that integrates GNNs into our sequence modeling pipeline. Within each block, input data is first processed by GNN modules to capture localized structural dependencies, followed by sequence modeling to encode long-range signal propagation. This integration enables end-to-end learning across RTL designs of varying structures.


In summary, our work makes the following key contributions:
\begin{enumerate} 
    \item We present RTL-Sequencer, the first to propose a sequence-based paradigm to RTL timing prediction. This paradigm enables scalable learning of long-range signal dependencies with linear computational complexity, offering explicit signal directionality and elimination of intra-layer restriction.
    \item We design four synergistic techniques customized to the sequence-based paradigm, namely sequence shuffling, bidirectional sequence modeling, differentiable sequence modeling, and a hybrid graph–sequence architecture.
    \item We conduct extensive experiments to demonstrate the superiority of RTL-Sequencer over state-of-the-art graph-based approaches. In particular, comprehensive ablation studies confirm the effectiveness of the proposed four techniques.
\end{enumerate}



\section{PROBLEM FORMULATION}

Our problem formulation follows prior works \cite{fang2024annotating,wangbridging}. Specifically, a digital circuit described at the RTL is represented as a directed graph $G=(V, E)$, where $V$ denotes the set of nodes corresponding to registers and combinational logic elements, and $E$ denotes the set of nodes corresponding to registers and combinational logic elements. This graph abstraction, often referred to as a Boolean Operator Graph (BOG) \cite{fang2024annotating,wangbridging}, encapsulates the structural and functional dependencies inherent in RTL designs. 

The primary objective is to predict the arrival time at each endpoint register node $n_{\mathrm{ep}}\in V_{\mathrm{ep}}$, denoted as $\mathrm{AT}_{\mathrm{pred}}(n_{\mathrm{ep}})$, directly from the RTL representation. Ground-truth arrival times $\mathrm{AT}_{\mathrm{label}}(n_{\mathrm{ep}})$ are obtained from sign-off static timing analysis (STA) performed after place-and-route. The prediction task is thus formulated as a supervised regression problem, where the goal is to minimize the discrepancy between $\mathrm{AT}_{\mathrm{pred}}(n_{\mathrm{ep}})$ and $\mathrm{AT}_{\mathrm{label}}(n_{\mathrm{ep}})$ across all register nodes. Formally, the optimization objective is expressed as:
\begin{equation}
    \min_{\theta} \sum\limits_{n_{\mathrm{ep}} \in V}{||\mathrm{AT_{pred}}(n_{\mathrm{ep}};\theta)-\mathrm{AT_{label}}(n_{\mathrm{ep}})||^2}
\end{equation}
where $V_{\mathrm{ep}}\subseteq V$ denotes the subset of endpoint register nodes, and $\theta$ represents the learnable parameters of the prediction model.

\begin{figure*}[!h]
    \centering
    \includegraphics[width=\linewidth]{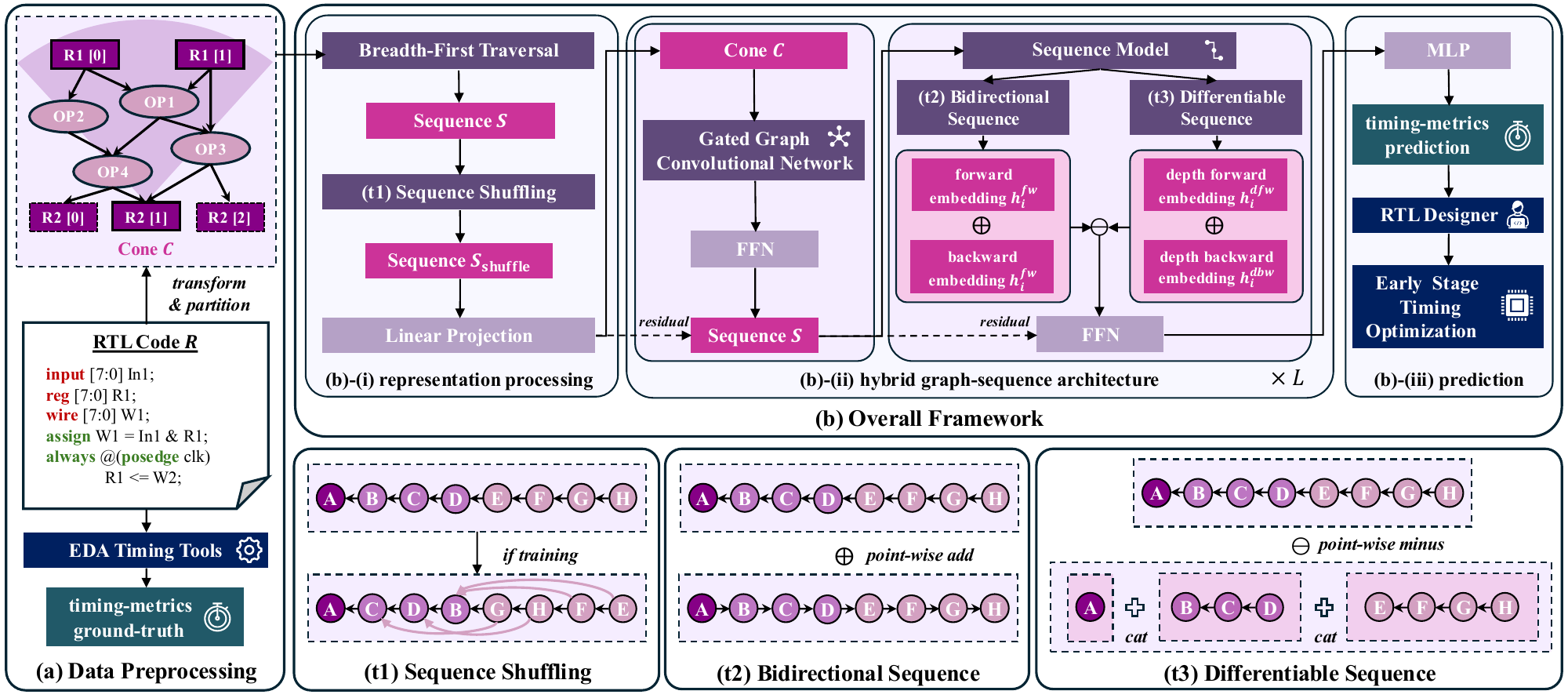}
    \vspace{-5mm}
    \caption{\textit{Overview of RTL-Sequencer.} (a) RTL data preprocessing includes extracting ground-truth timing metrics via commercial EDA tools and transforming RTL codes into BOGs. Subsequently, logic cones are extracted from the BOG. (b) Overall architecture of the proposed RTL-Sequencer framework, where the extracted logic cone from (a) is fed into a sequence modeling pipeline to predict timing metrics and facilitate neural network training. (t1) Sequence Shuffling, (t2) Bidirectional Sequence, and (t3) Differentiable Sequence illustrate key components of our sequence-centric paradigm, each implemented as submodules within the architecture shown in (b), and designed to improve model generalization across heterogeneous RTL designs.}
    \label{fig:method}
\end{figure*}

\section{PRELIMINARY: SEQUENCE MODELS}
In this section, we provide an overview of the background and motivation underlying modern linear sequence models \cite{gu2024mamba,dao2024transformers,sun2023retentive,peng2023rwkv,yang2024parallelizing}. In a general sequence model, the representation of a node is updated based on its current state and accumulated historical information. Notably, this historical context is losslessly encoded by preceding nodes while maintaining linear computational complexity. Such a formulation contrasts with attention-based architectures, in which the representation of each node is explicitly conditioned on all other nodes within the sequence, thereby incurring quadratic complexity.

Specifically, let an ordered sequence of nodes be denoted by $S=(n_1, n_2,..., n_i, ..., n_N)$. At each block, $x_i \in \mathbb{R}^{k}$ is the input feature vector for node $n_i$, $y_i \in \mathbb{R}^{k}$ be the corresponding output vector, and $h_{i-1} \in \mathbb{R}^{k}$ be the hidden state encoding the history from $n_1$ to $n_{i-1}$. The update formula for node representation is governed by:
\begin{align}
    h_i = &\  \texttt{Sequencer}(S, h_{i-1}, x_i) \label{eq2} \\
    y_i = &\  \texttt{Projection}(h_i)
\end{align}

Here, the $\texttt{Sequencer}$ captures the dependency between the current node $n_i$ and the history encapsulated in $h_{i-1}$ without strict node-to-node restrictions. The $\texttt{Projection}$ then refines channel-wise features of the updated hidden state $h_{i}$. The resulting $y_i$ is regarded as the subsequent input for the following block, thereby facilitating representation learning across successive network blocks.

\section{METHODOLOGY}

In this section, we present the methodology of RTL-Sequencer and depict it in Fig.~\ref{fig:method}. To establish clarity and emphasize our contributions, we first outline a basic solution for our framework in Sec.~\ref{sec:4_1}, detailing how to transform RTL designs into node sequences and the subsequent application of sequence models for representation learning. 
Building on this foundation, we introduce four synergistic techniques, including sequence shuffling in Sec.\ref{sec:4_3}, bidirectional sequence modeling in Sec.\ref{sec:4_4}, differentiable sequence modeling in Sec.\ref{sec:4_5}, and a hybrid graph-sequence architecture in Sec.\ref{sec:4_6}, which further customize sequence models to RTL timing prediction.

\subsection{Basic Solution} \label{sec:4_1}
To establish the sequence-based paradigm, we first outline a basic solution of RTL-Sequencer in this section. This involves transforming RTL designs into ordered node sequences and applying sequence models to predict timing behavior directly from these sequences. 

Initially, given that the timing behavior of an endpoint register is predominantly determined by its associated logic cone within the BOG, we adopt a cone-based partitioning strategy to isolate the cone, which is defined as the transitive fan-in subgraph rooted at the endpoint register.
Formally, for each endpoint register $\mathrm{ep} \in G$, we extract its logic cone $C\subseteq G$ via a backward traversal from $\mathrm{ep}$ to any input register, thereby capturing all upstream combinational and sequential elements that contribute to its signal arrival time. 

Subsequently, we adopt an inverse Breadth-First Traversal (BFT) initiated from the endpoint to convert the cone $C$ into the node sequence $S = \mathrm{BFT}(C)= (n_1, n_2,..., n_i, ..., n_N)$. In contrast to depth-first traversal, which may obscure timing dependencies due to its recursive exploration, BFT visits nodes in a level-wise order. This ordering preserves the hierarchical progression of signal propagation through the combinational gate and facilitates a more faithful representation of timing behaviour in the logic cone.

Such a node sequence $S$ is input to the proposed sequence models, denoted as \texttt{Sequencer}, in place of conventional graph-based approaches. For each node $n_i \in S$, the forward hidden state embedding $h^{\text{fw}}_{i}$ is computed by the sequence model, thereby capturing historical dependencies from preceding nodes in the sequence. Formally, the forward sequence embedding $h^{\text{fw}}_{i}$ of $n_i$ is defined as:
\begin{align}
    h^{\text{fw}}_{i} = &\ \texttt{Sequencer}(S,h_{i-1}, x_i) 
\end{align}
where $h_{i-1}$ denotes the hidden state propagated from the earlier nodes and $x_i$ represents the input feature vector associated with node $n_i$. In contrast to the basic formulation in Formula~\ref{eq2}, the final representation $h_i$ is not determined solely by $\texttt{Sequencer}(S,h_{i-1}, x_i)$; rather, it is augmented in subsequent stages through the integration of additional components that enrich the representational capacity.

In this section, we have introduced a basic solution for the RTL-Sequencer as a baseline. Nevertheless, the direct application of deterministic BFT and conventional sequence models proves sub-optimal, as it often discards critical topological information when topology-invariant cones are reduced to fixed, unidirectional and breadth-first sequences. To overcome these limitations, the subsequent sections present four synergistic techniques that are designed to further tailor sequence models to the challenging RTL timing prediction.

\subsection{Sequence Shuffling} \label{sec:4_3}

\begin{table}[!t]
\centering
\begin{tabular}{ccc}
\hline
Feature       & Type    & Channel \\ \cline{1-3} 
cell type     & one-hot & 12      \\
fan-out number & int     & 1      \\
fan-in number & int     & 1      \\
load capacitance & float & 1 \\
slew & float & 1 \\
node order in current BFT-level & int     & 1      \\
number of nodes in current BFT-level & int     & 1      \\
start flag of each BFT-level & one-hot     & 1      \\
end flag of each BFT-level  & one-hot    & 1      \\
order of parent node in next BFT-level & int & 1      \\ \hline
\end{tabular}
\vspace{1mm}
\caption{Overview of node representations as network inputs.}
\label{tab:node_feature}
\vspace{-10mm}
\end{table}

To address the information loss incurred when topology-invariant cones are converted into fixed node sequences, we introduce a stochastic shuffling mechanism. Specifically, each node ordering is randomized within each BFT depth level with 50\% probability, while preserving the cone topology's invariance. Hierarchical consistency is maintained by synchronizing permutations across child nodes in subsequent levels. This procedure serves as a form of data augmentation, exposing sequence models to diverse node orderings while retaining logical dependencies and timing closure, ultimately enhancing generalization across heterogeneous RTL designs.

For example, consider a cone sequence $S = (x_1, x_2, x_3, x_4, x_5)$, where $x_1$ connects to $\{x_2, x_3\}$, and $x_2$, $x_3$ further connect to $\{x_4\}$, $\{x_5\}$, respectively. $S_{\text{shuffle}} = (x_1, x_3, x_2, x_5, x_4)$ can be a possible shuffled variant, in which the position of $v_5$ is elevated relative to $v_4$, reflecting the direct connection between $x_3$ and $x_5$.

During inference, it is necessary to determine a sorting strategy consistent with the training procedure. To fully leverage the information flow inherent in sequence models, we employ a heuristic deterministic ordering that ranks nodes within each breadth‑first traversal (BFT) level according to their in‑degree and out‑degree statistics. Nodes exhibiting higher connectivity are positioned closer to the output register within each level, thereby enriching the contextual information available for timing prediction. This approach effectively integrates the advantages of stochastic data augmentation during training with the maximization of information utilization during inference, ultimately enhancing prediction accuracy across previously unseen RTL designs.

\subsection{Bidirectional Sequence Modeling} \label{sec:4_4}

\begin{table*}[!th]
\centering
\begin{tabular}{cc|ccc|ccc|ccc}
\hline
\multirow{2}{*}{Model} & \multirow{2}{*}{Type} & \multicolumn{3}{c|}{Arrival Time (AT)} & \multicolumn{3}{c|}{Worst Negative Slack (WNS)} & \multicolumn{3}{c}{Total Negative Slack (TNS)} \\ \cline{3-11} 
 & & $R$ & $R^2$ & MAPE & $R$& $R^2$ & MAPE & $R$& $R^2$ & MAPE \\ \hline
GCN \cite{kipf2016semi} & GNN & 0.81 & 0.66 & 27.38\% & 0.78 & 0.62 & 30.00\%  & 0.68 & 0.47 & 40.52\% \\
GAT \cite{velivckovic2017graph} & GNN & 0.83 & 0.69 & 25.23\%  & 0.81 & 0.65 &  27.46\%  & 0.69 & 0.48 &  39.11\%  \\
SG-Former \cite{wu2023sgformer} & GT  & 0.73 & 0.54 & 32.67\%  & 0.70 & 0.51 & 34.71\% & 0.66 & 0.43 & 46.80\%  \\ 
\hline
RTL-Timer \cite{fang2024annotating} & XGBoost & 0.85 & 0.72 & 23.55\% & 0.83 & 0.69 & 24.09\% & 0.70 & 0.50 & 37.10\% \\
RTLDistill \cite{wangbridging} & GNN-Propagation & 0.89 & 0.81 & 21.86\% & 0.87 & 0.74 & 23.11\% & 0.74 & 0.56  & 32.15\% \\
NUA-Timer \cite{wang2025pre} & GNN-Propagation & 0.90 & 0.81 & 21.16\% & 0.88 & 0.77 & 22.62\% & 0.76 & 0.59 & 29.45\%  \\
CircuitFusion \cite{fang2025circuitfusion} & GT & 0.91 & 0.83 & 20.29\% & 0.89 & 0.80 & 21.67\% & 0.82 & 0.69 & 26.16\%  \\
TF-Predictor \cite{cao2022tf} & Transformer & 0.70 & 0.50 & 35.74\% & 0.67 & 0.45 & 37.98\%  & 0.63 & 0.40 & 49.31\%  \\
\rowcolor{mycolor}
RTL-Sequencer & Sequencer & 0.92 & 0.85 & 17.24\% & 0.91 & 0.83 & 17.66\% & 0.88 & 0.77 & 23.92\% \\ \hline
\end{tabular}
\vspace{1mm}
\caption{Comparison between classic graph-based solutions, SOTA in the community, and RTL-Sequencer on timing prediction.}
\label{tab:main}
\vspace{-4.5mm}
\end{table*}

\begin{table*}[!th]
\centering
\begin{tabular}{c|ccc|ccc|ccc}
\hline
\multirow{2}{*}{Model} & \multicolumn{3}{c|}{Arrival Time (AT)} & \multicolumn{3}{c|}{Worst Negative Slack (WNS)} & \multicolumn{3}{c}{Total Negative Slack (TNS)} \\ \cline{2-10} 
 & $\Delta$$R$        & $\Delta R^2$       & $\Delta$MAPE       & $\Delta$$R$& $\Delta R^2$          & $\Delta$MAPE          & $\Delta$$R$& $\Delta R^2$          & $\Delta$MAPE          \\ \hline
\rowcolor{mycolor}
RTL-Sequencer & 0.92 & 0.85 & 17.24\% & 0.91 & 0.83 & 17.66\% & 0.88 & 0.76 & 23.92\% \\ \hline
Basic Solution & -0.12 & -0.22 & +12.23\% & -0.13 & -0.21 & +12.92\% & -0.19 & -0.26 & +16.39\% \\
w/o Stochastic Shuffling & -0.09 & -0.14 & +8.23\% & -0.11 & -0.18 & +9.26\% & -0.16 & -0.22 & +11.69\% \\
w/o Bidirectional Sequence Modeling & -0.04 & -0.09 & +5.01\% & -0.04 & -0.05 & +5.73\% & -0.11 & -0.15 & +7.20\% \\
w/o Differentiable Sequence Modeling & -0.01 & -0.02 & +2.37\% & -0.02 & -0.04 &  +2.78\% & -0.06 & -0.08 & +3.46\%  \\
w/o Hybrid Architecture & -0.02 & -0.02 & +3.22\% &  -0.03  &  -0.05   &  +3.93\%  & -0.08   &  -0.11   &  +5.18\%  \\ \hline
\end{tabular}
\vspace{1mm}
\caption{Ablation study of components in RTL-Sequencer compared to the full model.}
\label{tab:ablation_components}
\vspace{-4.5mm}
\end{table*}

In the baseline model described in Sec.~\ref{sec:4_1}, a single sequence model is employed to capture the forward sequence embedding $h^{\text{fw}}_{i}$ for node $n_i$. This formulation, however, constrains information propagation to a unidirectional flow, thereby preventing later nodes from influencing earlier representations and limiting the depth of sequencer blocks. Such a restriction is also misaligned with practical design processes, as both Static Timing Analysis (STA) and logical synthesis inherently operate not in unidirectional contexts.

To overcome this limitation, we propose a bidirectional sequence modeling to augment the forward sequence $S$ by its reversed counterpart, denoted as $S_{\mathrm{flip}}=(n_N,...,n_i, ..., n_2, n_1)$. A reversed sequence model is instantiated to process $S_{\mathrm{flip}}$, thereby facilitating backward information flow. For each node $n_i$, the backward hidden state $h^{\text{bw}}_{i}$ is computed by aggregating the outputs of backward sequencers:
\begin{align}
    h^{\text{bw}}_{i} = &\ \texttt{Sequencer}(S_{\mathrm{flip}}, h_{i+1}, x_i)
\end{align}

This dual-path formulation markedly enhances representational expressiveness by embedding each node within both its antecedent and subsequent nodes, thereby enabling the model to more effectively capture timing-critical dependencies across deep logics.

\subsection{Differentiable Sequence Modeling} \label{sec:4_5}
While breadth-first traversal avoids backtracking compared to depth-first traversal, it frequently places unrelated nodes at the same BFT depth, thereby placing them adjacently within the resulting node sequence. In sequence models, this adjacency causes each node’s representation to be inevitably influenced by its immediate depth-level neighbors. As a result, sequence models tend to overemphasize local depth-specific structures while neglecting long-range dependencies along critical timing paths.

To address this limitation, we propose a differentiable mechanism that adaptively reduces irrelevant depth-specific context while accentuating global timing behaviors. Our design is inspired by recent advances in differentiable sequence modeling that employ embedding subtraction across entire sequences \cite{ye2024differential}. Distinct from these approaches, we introduce a sequence model tailored to operate on depth-specific subsequences rather than the full sequence. By subtracting depth-specific node embeddings from those obtained via the global sequence model, we derive depth-irrelevant embeddings that more faithfully preserve global timing dependencies.

For instance, consider the sequence $S = (n_1, n_2, n_3, n_4, n_5)$, where $n_1$ has children $\{n_2, n_3\}$, and $n_2$, $n_3$ connect to $\{n_4\}$, $\{n_5\}$, respectively. The corresponding BFT-level sequences $D_i$ are defined as $D_1=(N_1)$, $D_2=D_3=(n_2,n_3)$, $D_4=D_5=(n_4,n_5)$. For each node $n_i$, we first compute the depth-specific embeddings $h^{\text{dfw}}_{i}$ and $h^{\text{dbw}}_{i}$ from the forward and backward sequences as follows:
\begin{align}
      h^{\text{dfw}}_{i} &\ =  \texttt{Sequencer}(D_{i},h_{i-1}, x_i) \\
     h^{\text{dbw}}_{i}  &\ = \texttt{Sequencer}(D_{i\_\text{flip}}, h_{i+1}, x_i)
\end{align}

We further incorporate a self-gating mechanism to adaptively regulate the strength of the differential operation. Specifically, we introduce the learnable coefficient $\lambda$ that quantifies the influence of depth-specific context. Owing to the $\texttt{Sigmoid}$ activation function, $\lambda$ is constrained to the interval $[0,1]$. Ideally, $\lambda \approx 0$ indicates that the node embedding predominantly attends to global timing behavior, whereas  $\lambda \approx 1$ reflects a stronger emphasis on depth-local structural information, allowing the representation to self-adjust through learning. The overall formulation is expressed as:
\begin{align}
    \lambda &\ = \texttt{Sigmoid} (\texttt{MLP}(h^{\text{dfw}}_{i}+h^{\text{dbw}}_{i})) \\
     h_i  &\ =  h^{\text{fw}}_{i} + h^{\text{bw}}_{i} -\lambda(h^{\text{dfw}}_{i} + h^{\text{dbw}}_{i})
\end{align}
Here, $h^{\text{fw}}_{i}$ and $ h^{\text{bw}}_{i}$ denote the forward and backward global embeddings obtained from Sec.~\ref{sec:4_1} and Sec.\ref{sec:4_3}, respectively. This differentiable strategy (i.e., global minus depth-local) ensures that node representations remain aligned with the global timing behavior rather than adaptively aligning with depth-specific structure.

\subsection{Hybrid Architecture} \label{sec:4_6}
While sequence models are effective in capturing long-range and deep dependencies, they can fail to localize and broader structural information surrounding nodes within design logic. To overcome this limitation, we propose a hybrid graph–sequence architecture that integrates the complementary strengths of GNNs for local context aggregation with the capacity of sequence models to capture global and deep temporal dependencies.

As illustrated in Fig.~\ref{fig:method}-(b), the proposed pipeline begins by applying BFT combined with sequence shuffling to the logic cone, thereby generating the initial sequence representation. This feature channel of sequence is subsequently projected into a higher-dimensional space through a linear transformation. Following this representation processing, a GNN is employed to capture broader local structural dependencies surrounding each node in the cone. The cone is subsequently converted back to an ordered sequence and processed by our sequence models, which are augmented with bidirectional and differential sequence modeling, to capture the long-range and deep timing dependencies. To further enrich channel-wise feature representations, feed-forward networks (FFNs) are appended to both the GNN and sequence models, following the design principles of Transformer-based architectures \cite{vaswani2017attention}. Crucially, all graph-to-sequence and sequence-to-graph transitions are implemented as differentiable operations, thereby enabling seamless end-to-end optimization through gradient descent.

\vspace{-2mm}

\section{EXPERIMENTS}
\subsection{Configurations}
\textbf{Dataset.} We conduct experiments on 21 open-source RTL designs following RTL-Timer~\cite{fang2024annotating}, employing 5-fold cross-validation. To ensure generalization, training and evaluation sets are strictly partitioned by designs. The dataset encompasses diverse mainstream hardware description languages, with design sizes ranging from 6K to 510K gates. For dataset generation, we utilize Synopsys Design Compiler for logic synthesis and Cadence Innovus for physical implementation, both targeting the NanGate45nm process design kit. Static timing analysis is performed using Synopsys PrimeTime to extract ground-truth timing metrics.

\textbf{Evaluation Metrics.} We evaluate model performance on three critical timing metrics: Arrival Time (AT), Worst Negative Slack (WNS), and Total Negative Slack (TNS). The following statistical measures are employed: \textbf{Pearson Correlation Coefficient ($R$)}: Quantifies the linear correlation between predicted and actual timing values. \textbf{Coefficient of Determination ($R^2$)}: Measures the proportion of variance in ground truth explained by the model. \textbf{Mean Absolute Percentage Error (MAPE)}: Captures the relative prediction error, with lower values indicating higher accuracy.

\textbf{Implementation Details.} We adopt the GatedGCN \cite{li2015gated} and Mamba-2 \cite{dao2024transformers} for our hybrid graph-sequence architecture. All models are trained on a cluster of $8 \times$ NVIDIA RTX 4090D using PyTorch Lightning, PyTorch Geometric, and Scikit-learn. We adopt the Adam optimizer with an initial learning rate of $1 \times 10^{-3}$ and a per-GPU batch size of 8. For the model hyper-parameters, the channel number of the hidden state is 32, the number of sequencer blocks $L$ is 3, and shuffling probability applied to each training sample is 50\%.

\textbf{Baseline.} We benchmark our framework against three representative graph-based approaches, including GCN \cite{kipf2016semi}, GAT \cite{velivckovic2017graph}, and SG-Former \cite{wu2023sgformer}; as well as several SOTA methods widely adopted in the community, including RTL-Timer \cite{fang2024annotating}, RTLDistill \cite{wangbridging}, NUA-Timer \cite{wang2025pre}, CircuitFusion \cite{fang2025circuitfusion}, and TF-Predictor \cite{cao2022tf}. For consistency and fairness, all baselines are re-implemented within our setting, deliberately excluding auxiliary features derived from post-synthesis stages (e.g., layout information). This design choice highlights the contribution of architectural innovations to RTL timing prediction.

\vspace{-2mm}

\subsection{Results}

\begin{figure}[!t]
    \centering
    \includegraphics[width=\linewidth]{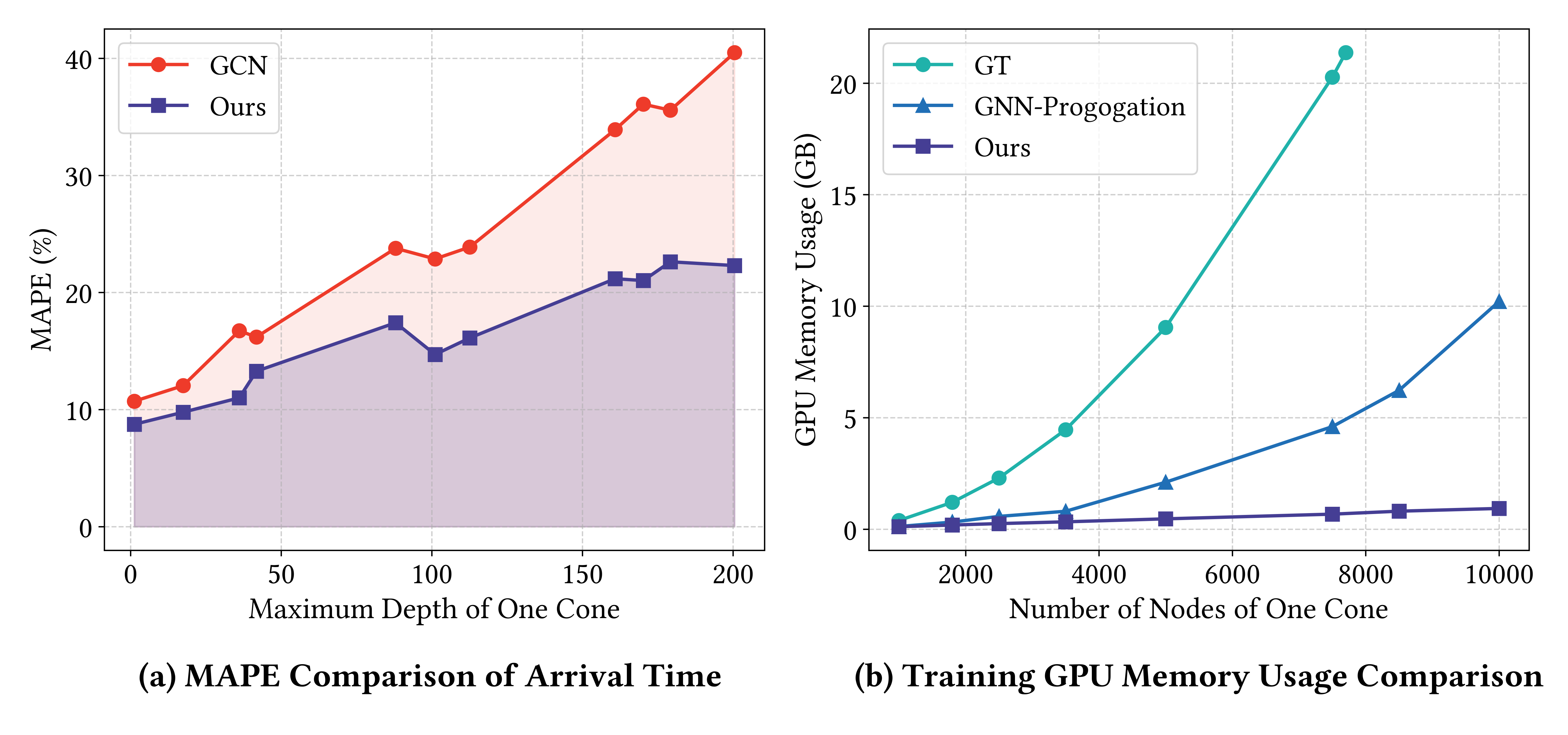}
    \vspace{-7mm}
    \caption{Scalability comparison among GCN, GT, GNN-Propagation, and RTL-Sequencer demonstrates that our superior scalability in handling deeper cones with more nodes.}
    \label{fig:3_structure}
    \vspace{-4.3mm}
\end{figure}

As presented in Table~\ref{tab:main}, RTL-Sequencer consistently surpasses prior approaches across all evaluated timing metrics, thereby highlighting the efficacy of its architectural innovations. Specifically, the framework achieves the lowest AT MAPE of 17.24\%, accompanied by the highest correlation scores, with $R=0.92$ and $R^2=0.85$
. In addition, RTL-Sequencer demonstrates marked superiority on TNS, attaining a MAPE of 23.92\%, whereas the second-best method, CircuitFusion \cite{fang2025circuitfusion}, records 29.45\%. This performance gap can be attributed to the absence of an explicit representation of signal propagation in attention mechanism of CircuitFusion, which only models node-to-node relationships. Although both CircuitFusion and SG-Former \cite{wu2023sgformer} adopt GT-based architectures, SG-Former yields a relatively poor AT MAPE of 32.67\%, due to its reliance on a single block combining a GT with a GNN. RTLDistill \cite{wangbridging} and NUA-Timer \cite{wang2025pre} are hindered by exponential accumulation of errors across cone layers, thus are outperformed by RTL-Sequencer. Notably, TF-Predictor \cite{cao2022tf} exhibits the weakest performance among all baselines, with an AT MAPE of 35.71\%. Despite its claim as a sequence-based model, TF-Predictor relies on a naive Transformer architecture that focuses exclusively on specified timing paths. This design restricts its applicability to RTL stages and ultimately renders it ineffective for modeling meaningful representations of logic cones.

Moreover, Fig.~\ref{fig:3_structure} illustrates two fundamental limitations inherent in prior graph-based approaches. As shown in Fig.~\ref{fig:3_structure}(a), the performance of GCN deteriorates more compared to RTL-Sequencer as logic depth increases, highlighting its restricted capacity to effectively capture deep combinational structures. In Fig.~\ref{fig:3_structure}(b), GT exhibits quadratic memory overhead of the node count when the batch size is set to one, rendering it impractical for large-scale designs. Furthermore, by incrementally adding 100 nodes per layer within the cone, we demonstrate that the computational complexity of GNN-Propagation scales quadratically with cone depth. In contrast, RTL-Sequencer maintains linear memory complexity, thereby enabling scalable training across deep and complex designs.

\subsection{Ablation Studies}

To evaluate the individual contributions of the proposed architectural innovations, we conducted a comprehensive ablation study, the results of which are summarized in Table~\ref{tab:ablation_components}. The baseline solution can achieve accuracy comparable to graph-based approaches, thereby underscoring the promise of the sequence-based paradigm. Moreover, each component contributes measurable improvements in predictive performance, highlighting the synergistic nature of the overall framework. Among these, the sequence shuffling mechanism delivers the most pronounced gains in accuracy. This improvement arises from its function as a structural data augmentation strategy: by introducing stochastic permutations within BFT levels, the model is exposed to a wider range of topological variations while maintaining logical dependencies. Such diversity is able to mitigate overfitting to the deterministic node ordering, thereby enhancing generalization across heterogeneous RTL designs.

\begin{figure}[!t]
    \centering
    \includegraphics[width=\linewidth]{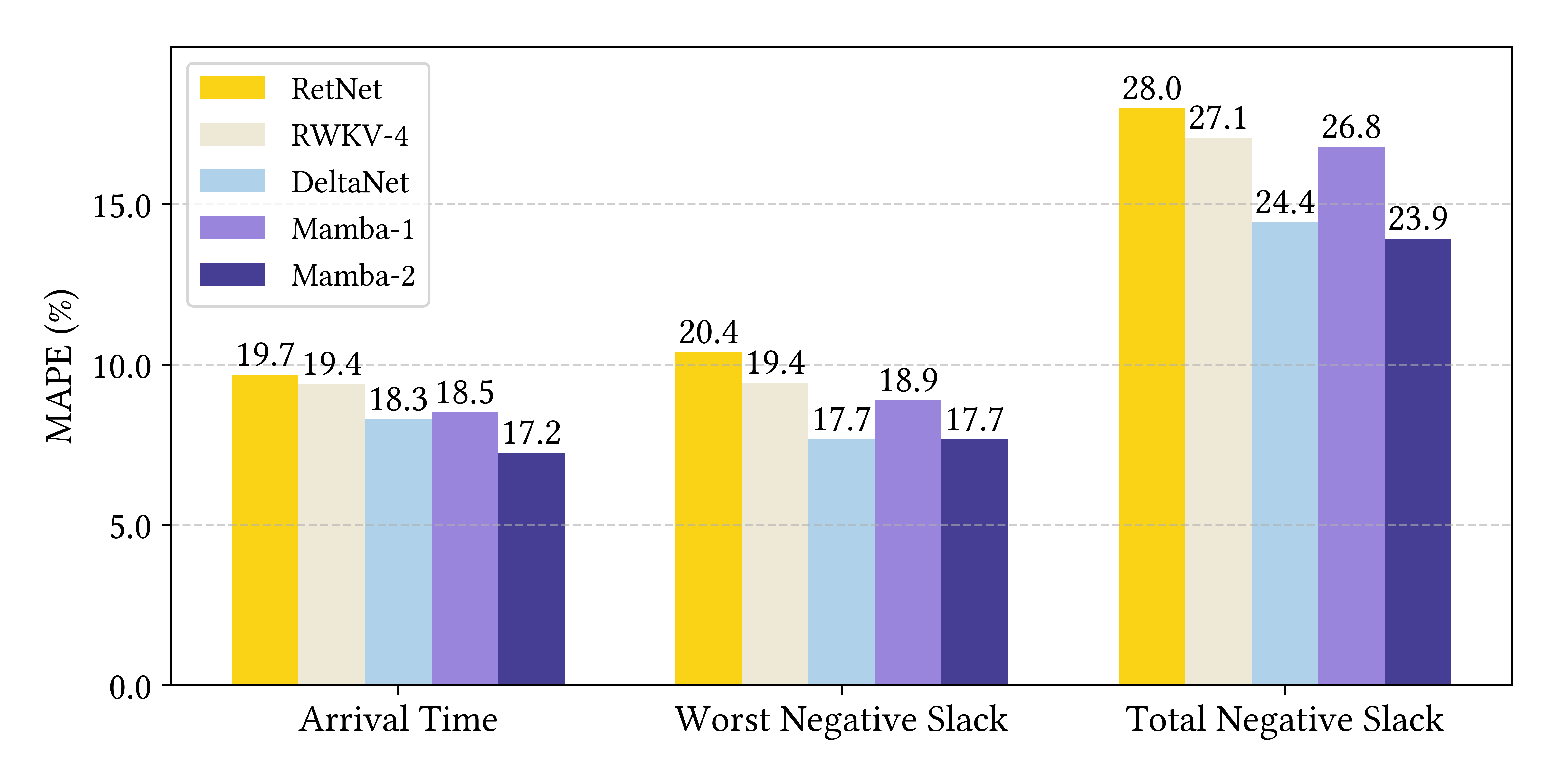}
    \vspace{-7mm}
    \caption{Comparison across sequence models. Mamba-2 achieves the lowest MAPE across all metrics. Notably, all sequence models surpass graph-based solutions.}
    \label{fig:ablation_sequence_models}
    \vspace{-4.3mm}
\end{figure}

To rigorously evaluate the effectiveness of the proposed sequence modeling paradigm for RTL timing prediction, we conducted a comprehensive comparative study across four representative sequence architectures. As illustrated in Fig.~\ref{fig:ablation_sequence_models}, the Mamba-2 model consistently delivers superior performance across all timing metrics, outperforming alternative sequence-based approaches. Notably, all sequence models substantially exceed traditional graph-based solutions, including GNN and GT variants, in both predictive accuracy and scalability. This consistent superiority highlights a fundamental paradigm shift from isotropic graph representations to directional sequence modeling, thereby substantiating the core hypothesis that logic cone linearization enables more expressive and scalable timing prediction. Collectively, these findings underscore the promise of sequence-driven frameworks as a robust and scalable alternative to graph-based methodologies for early-stage design automation.

\section{CONCLUSION}
This work introduces RTL-Sequencer, a novel sequence-based paradigm for scalable RTL timing prediction. By converting logic cones into ordered node sequences and employing advanced sequence modeling techniques, the framework achieves global receptive fields with linear computational complexity while explicitly preserving signal directionality. The integration of sequence shuffling, bidirectional sequence modeling, differentiable sequence modeling, and a hybrid graph–sequence architecture collectively enhances predictive accuracy and scalability, consistently outperforming SOTA graph-based approaches. Comprehensive experiments and ablation studies substantiate the effectiveness of both the overall paradigm and its individual components. Taken together, these results demonstrate the potential of RTL-Sequencer to advance early-stage timing estimation in digital design automation significantly.

\section{ACKNOWLEDGEMENT}
This work is supported by Hong Kong Research Grants Council (RGC) CRF-YCRG C6003-24Y, GRF 16200724, and T46-415/25-R. It was partially conducted by ACCESS – AI Chip Center for Emerging Smart Systems, supported by the InnoHK initiative of the Innovation and Technology Commission of the Hong Kong Special Administrative Region Government.

\newpage
\bibliographystyle{ACM-Reference-Format}
\bibliography{acmart}
\end{document}